\documentclass[9pt,twocolumn,twoside,unsortedaddress]{revtex4-2}

\usepackage{multirow}
\usepackage{graphicx}
\usepackage[colorlinks=true, linkcolor=blue, urlcolor=blue, citecolor=blue]{hyperref}

\usepackage[font=small,labelfont=bf,labelsep=colon]{caption}

\usepackage{amsmath,amssymb}
\usepackage{enumitem}

\def\l{\left}
\def\r{\right}

\begin{document}

\title{Chimeric Forecasting: Blending Human Judgment and Computational Methods for Improved, Real-time Forecasts of Influenza Hospitalizations}

\author{Thomas McAndrew}
\affiliation{Department of Community and Population Health, College of Health, Lehigh University, Bethlehem, Pennsylvania, United States of America}

\author{Mark Lechmanik}
\affiliation{P.C. Rossin College of Engineering \& Applied Science, Lehigh University, Bethlehem PA, United States of America}

\author{Erin~N.~Hulland}
\affiliation{Boston Children’s Hospital, Boston, MA, United States of America}
\affiliation{Harvard Medical School, Boston, MA, United States of America}

\author{Shaun Truelove}
\affiliation{Department of Epidemiology, Johns Hopkins Bloomberg School of Public Health, Baltimore, Maryland, USA}

\author{Mark Ilodigwe}
\affiliation{Department of Community and Population Health, College of Health, Lehigh University, Bethlehem, Pennsylvania, United States of America}

\author{Maimuna Majumder}
\affiliation{Boston Children’s Hospital, Boston, MA, United States of America}
\affiliation{Harvard Medical School, Boston, MA, United States of America}


\begin{abstract}
Infectious disease forecasts can reduce mortality and morbidity by supporting evidence-based public health decision making. Most epidemic models train on surveillance and structured data (e.g. weather, mobility, media), missing contextual information about the epidemic. Human judgment forecasts are novel data, asking humans to generate forecasts based on surveillance data and contextual information. Our primary hypothesis is that an epidemic model trained on surveillance plus human judgment forecasts (a chimeric model) can produce more accurate long-term forecasts of incident hospitalizations compared to a control model trained only on surveillance. Humans have a finite amount of cognitive energy to forecast, limiting them to forecast a small number of states. Our secondary hypothesis is that a model can map human judgment forecasts from a small number of states to all states with similar performance. For the 2023/24 season, we collected weekly incident influenza hospitalizations for all US states, and 696 human judgment forecasts of peak epidemic week and the maximum number of hospitalizations (peak intensity) for ten of the most populous states. We found a chimeric model outperformed a control model on long-term forecasts. Compared to human judgment, a chimeric model produced forecasts of peak epidemic week and peak intensity with similar or improved performance. Forecasts of peak epidemic week and peak intensity for the ten states where humans input forecasts vs a model that extended these forecasts to all states showed similar performance to one another. Our results suggest human judgment forecasts are a viable data source that can improve infectious disease forecasts and support public health decisions.
\end{abstract}

\maketitle

\section{Introduction}

Infectious disease forecasts give advanced warning to public health officials---those working at federal, state, or more local jurisdictions---about potential changes in transmission and burden associated with a pathogen~\cite{mcandrew2021adaptively,ray2018prediction,osthus2019dynamic,osthus2017forecasting,osthus2021multiscale,gibson2021improving,ben2019forecasting,riley2017identifying,turtle2021accurate,lemaitre2024flepimop}.
Forecasts have been successful in two broad categories: (1) warning about increases in transmission for small time horizons~(up to 4 weeks)~\cite{reich2019collaborative,reich2019accuracy,reich2021zoltar,mathis2024title,mcgowan2019collaborative} and (2) understanding how interventions can mitigate infectious disease impact~\cite{loo2024us,borchering2023public}.  
As a result of their importance, forecast hubs that collect, combine, and communicate ensemble forecasts have been formed~\cite{reich2022collaborative,cramer2022united,ray2020ensemble,howerton2023evaluation,cramer2022evaluation}. 
Within the context of seasonal influenza, which causes an average of 400k hospitalizations and 30k deaths annually in the US, forecasts of incident hospitalizations have been used to inform national public health efforts~\cite{lutz2019applying,cdcflu}.

The challenge that all forecasts must address is the long lag between when a policy change is decided, implemented, and an impact is observed~\cite{dey2021lag}.
Because of this lag, we argue that long-term forecasts~(forecasts 5 weeks or longer) of quantities like incident hospitalizations  have more importance when compared to short-term forecasts.

Incident hospitalization is a direct measure of the burden and severity of the flu season~\cite{rolfes2018annual,lafond2021global,giacchetta2022burden}.
Weekly incident hospitalizations are reported during the typical influenza season in the northern hemisphere, a season that typically starts on epidemic week~(epiweek) 40 of year $Y$ to epiweek 20 of year $Y+1$.
Two important attributes of the season that can guide policy are peak epidemic week and peak intensity. 
In this work, peak intensity is defined as the maximum number of reported hospitalizations over the course of the influenza season and the peak epidemic week is defined as the week at which the peak intensity was observed.

To generate forecasts of future incident hospitalizations~(including the peak epidemic week and peak intensity), the majority of contemporary forecasting techniques train on structured data~\cite{mathis2024title}. 
A novel approach to forecasting involves augmenting traditional models to take as input structured data like public health surveillance data (e.g., historical cases, hospitalizations, and deaths) as well as forecasts generated by humans---i.e., human judgment forecasts~\cite{farrow2017human,bosse2023human,bosse2022comparing,mcandrew2021aggregating,mcandrew2022chimeric,mcandrew2022early,mcandrew2024assessing,mcandrew2022expert,codi2022aggregating,braun2022crowdsourced,mcandrew2022aggregating,mcandrew2024chimeric}.
Traditional computational models trained on surveillance data typically also include auxiliary data sources like weather patterns~\cite{roussel2016quantifying,shaman2009absolute,shaman2017use,tamerius2013environmental}, genetic or viral typing~\cite{wolf2010projection,kandula2017type,hill2019seasonal}, social media~\cite{santillana2015combining,lee2017forecasting,alessa2018review}, and mobility patterns~\cite{venkatramanan2021forecasting}.  
In contrast, a human judgment forecast inputs a set of individual forecasts generated by multiple humans and outputs a single, unified forecast based on their collective intuition~\cite{farrow2017human,mcandrew2022early,mcandrew2024assessing,mcandrew2022expert,codi2022aggregating,mcandrew2022aggregating,bosse2022comparing,bosse2023human}.
An ensemble that combines computational and human forecasts or a model that trains on a combination of surveillance data and human forecasts---called \textit{chimeric} models---have been shown to have strong performance compared to alternative approaches~\cite{braun2022crowdsourced,mcandrew2024chimeric}.

The advantages to using human judgment forecasts are that they have been shown to be as accurate as statistical forecasts, can provide forecasts when data is sparse, and in the context of infectious disease forecasting, are flexible enough to quickly address changing public health needs~\cite{mcandrew2022expert,mcandrew2022early,mcandrew2024assessing,codi2022aggregating,mcandrew2022aggregating,mcandrew2022chimeric}.
There also exist several challenges to human judgment forecasts. 
Humans have been shown to rely on fast, heuristic processes; inject their own biases when solving problems, and find patterns in data when none truly exist~\cite{tversky1974judgment,goodwin2022forecasting,ellerby2017effects}.   

However, previously we have shown that---given synthetic surveillance data meant to mimic influenza---a model trained on surveillance data and human judgment forecasts of peak epidemic week~(epiweek) and peak intensity can outperform a control model trained only on the former~\cite{mcandrew2024chimeric}.

Thus, here we hypothesize that human judgment forecasts of peak epiweek and peak intensity can augment real surveillance data to produce better performing forecasts on average across all 50 states in the US during the influenza season.
We propose a discrete-time Kermack McKendrick model as the platform to combine both sources of data~\cite{diekmann2021discrete}.

Over a typical influenza season, we would have to ask for weekly forecasts over all 50 states over the course of the 32 week season~\cite{mathis2024title}.
Though machines can be programmed to produce forecasts for several locations and several time points, over any number of weeks, humans lack the cognitive energy to produce a large number of forecasts.
Instead, we pose a secondary goal to maximize the use of the limited cognitive energy that humans can apply to forecasting problems. In this work, we propose an additional machine model to extend human judgment forecasts from a small set of states to all states in the US.

To support our two hypotheses, we collected novel human judgment ensemble forecasts of peak epidemic week and peak intensity from the Metaculus platform from October 18, 2023 to March 2nd, 2024, and for the most populace state in each of the 10 Health and Human Service regions~\cite{metacflu2324}.
Metaculus is a forecasting and crowdsourcing platform that hosts `challenges' to focus efforts on specific tasks, such as building forecasts of seasonal influenza to support public health decision making.
As part of the Metaculus `FluSight Challenge', $25$ humans produced on average $7$ forecasts for each state, resulting in a mean 113 forecasts per question~(peak epiweek and peak intensity). 

\section{Results}
We present two major results:~(Result~\ref{betterpeak})~a chimeric seasonal influenza model led to improved forecasts of peak epiweek and peak intensity compared to a human judgment only~(HJ-only) ensemble~(i.e. excluding computational models); and (Result~\ref{betterlong})~when compared to a control model trained only on surveillance data, a chimeric model produced improved long-term forecasts. 
These results show that a chimeric model can retain and improve upon the quality of human judgment forecasts and, conversely, that human judgment can increase the predictive performance of more traditional epidemic models. 

To support our secondary hypothesis that a machine model can extend human judgment forecasts~(Result~\ref{result3}), we compare human judgment forecasts of peak epidemic week and peak intensity for states that were presented to individuals on Metaculus vs forecasts of peak epidemic week and peak intensity of states that were not presented to humans, but instead were generated by a proposed model~(Materials and Methods~\ref{hjextend}) that included human judgment.

For clarity of presentation, the control model and chimeric epidemic models have the same structure and make the same set of assumptions~(See Materials and Methods sections \ref{control_model} 
 and \ref{chimeric_model} for approach). 
The difference between these models is that the control model is trained on incident hospitalizations generated by surveillance, while the chimeric is trained on the same surveillance data plus human judgment forecasts of the peak epidemic week and peak intensity. The HJ-only ensemble presented below is a combination of human judgment forecasts only, the same forecasts that are included as data in the chimeric model~(See \ref{hjbuild} for methodology).

Our metric for performance is the relative weighted interval score or RWIS~\cite{bracher2021evaluating}.
We use this score to compare the chimeric model vs control and, separately, the HJ-only ensemble vs control.
This is a standard metric used in the forecasting community where negative values indicate the evaluated model outperformed the reference or control model, and 
smaller values indicate better forecast performance~\cite{mathis2024title}. 

\subsection{Improved peak prediction}
\label{betterpeak}

Chimeric model forecasts of peak epidemic week and peak intensity on average outperformed the control model forecasts before, at, and after peak hospitalization~(negative RWIS values). 
Before peak hospitalization, chimeric model forecasts of peak epidemic week and peak intensity performed similarly to HJ-only ensemble forecasts~(Table~\ref{tab1.RWIS}). 

\begin{table}[ht!]
    \centering
    \begin{tabular}{r|lll}
                \hline
                    Model  &  Before peak & At peak & After peak \\ 
                \hline

    Peak epidemic intensity     & \\
       \hspace{2em} Chimeric    & -0.86~(0.01) & -0.97~(0.01) & -0.47~(0.60) \\
       \hspace{2em} HJ-only     & -0.94~(0.01) & -0.85~(0.01) &  \hspace{2px}4.26~(0.01)  \\
       \hspace{2em} pvalue      & 0.22  & 0.194 & ${<}$ 0.01\\
       
    Peak epidemic week\\
       \hspace{2em} Chimeric    & -0.66~(0.01) & -0.59~(0.01) & -0.25~(0.01) \\
       \hspace{2em} HJ-only     & -0.54~(0.01) & -0.77~(0.01) & -0.91~(0.01) \\ 
       \hspace{2em} pvalue      & 0.31  & 0.01 &0.01\\
   
       \hline
    \end{tabular}
    \caption{
    Comparison of chimeric and human judgment only~(HJ-only) at three time points: 4 weeks before the peak, on the peak, and 4 weeks after the peak. 
    The mean relative weighted interval score~(RWIS) for forecasts of peak intensity and peak epidemic week is reported with the control model as reference.   
    Smaller values indicate better performance.
    A negative value indicates better performance when compared to control.
    Pvalue below each pair of scores compares RWIS between the chimeric and human judgment only models. Pvalue in parentheses compares performance between the presented models and control. \label{tab1.RWIS}}
\end{table}

\begin{figure}
    \centering
    \includegraphics[width=1.0\linewidth]{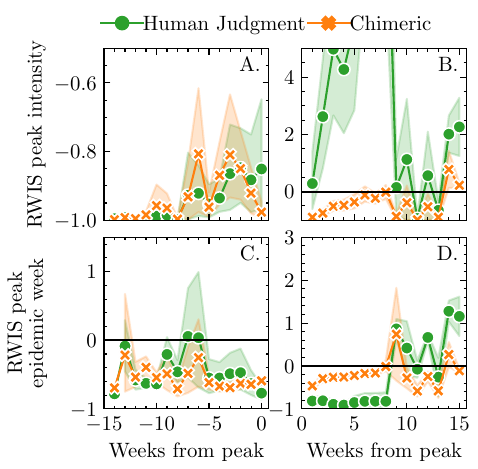}
    \caption{Average relative WIS~(RWIS) scores and 95\% confidence intervals for a chimeric model~(orange; trained on surveillance plus human judgment) and human judgment only ensemble~(green) compared to a control model~(trained on surveillance only) stratified by peak intensity and peak epidemic week, and before vs after observed peak incident hospitalizations.
    Positive RWIS values indicate worsened performance compared to control and negative RWIS values indicate improved performance compared to control. \label{fig.peakscores}}
\end{figure}

Forecast performance of peak epidemic intensity for the chimeric and HJ-only ensemble had similar trajectories that increased~(worsened) for forecasts fifteen to five weeks before observed ground truth peak hospitalizations~(Fig.~\ref{fig.peakscores}A.). 
From five weeks up until the peak, HJ-only ensemble forecasts of peak intensity continued to worsen but chimeric forecasts began to improve.  
Just after the peak was observed, HJ-only ensemble forecasts worsened while chimeric model forecasts of peak intensity continued to improve~(Fig.~\ref{fig.peakscores}B.).

Forecast performance of peak epidemic week from the start of season until the peak were similar for chimeric and HJ-only ensembles, exhibiting a constant improvement~(negative RWIS) versus control~(Fig.~\ref{fig.peakscores}C.). 
For one to eight weeks after the peak, the HJ-only ensemble was improved compared to the chimeric model until both models show varied performance when compared to control~(Fig.~\ref{fig.peakscores}D.).
\subsection{Long-term forecast improvement}
\label{betterlong}
\begin{figure*}[ht!]
    \centering
    \includegraphics[width=1.0\textwidth]{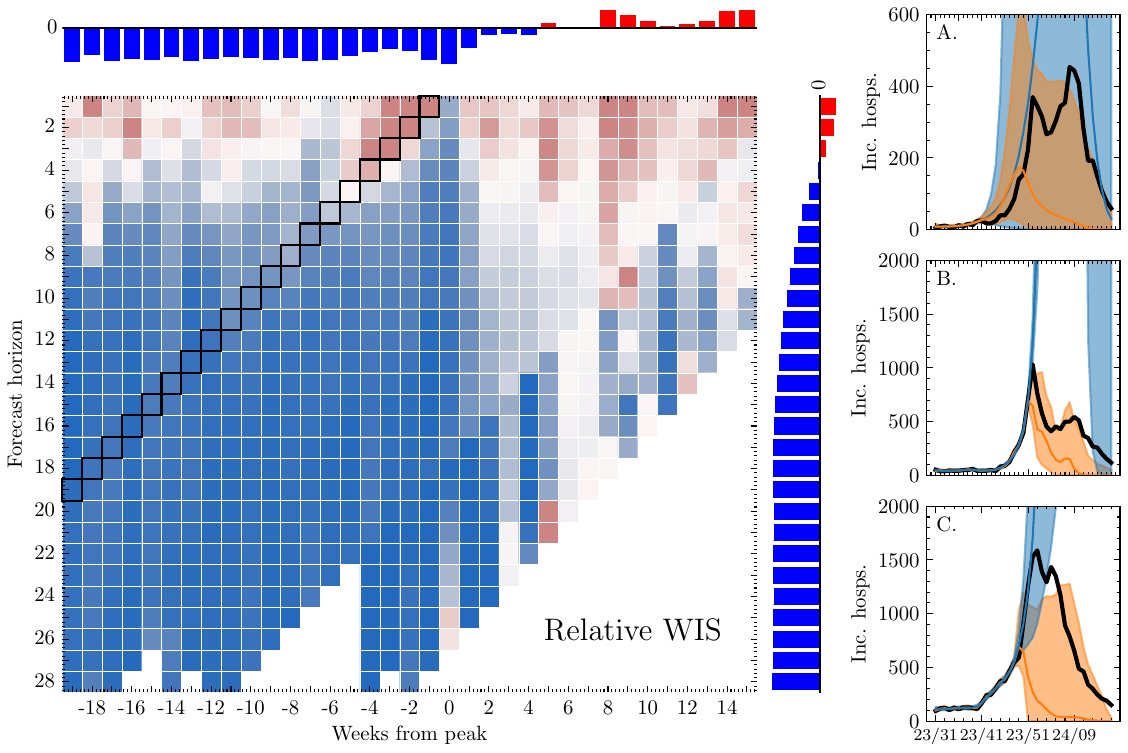}
    \caption{ Heatmap denoting the mean Relative WIS---where red indicates values above zero, or when the control is more accurate and where blue indicates values below zero, or when the chimeric model is better---for forecasts of incident hospitalizations stratified by the number of weeks until the peak and forecast horizon. 
    The black highlighted boxes indicate predictions of the peak number of hospitalizations.
    The top~(right) bar is the mean Relative WIS stratified by weeks until peak~(forecast horizon).
    The right column of forecasts, represented as a median and 95\% prediction interval, presents three examples of how a chimeric forecast~(orange) behaves compared to control~(blue). 
    (A.)~Forecasts produced on 2023-10-14 for Missouri
    (B.)~Forecasts produced on 2023-12-23 for Pennsylvania, and 
    (C.)~Forecasts produced on 2023-12-23 for Texas.  
    \label{fig.scoreweek} }
\end{figure*}

Compared to a control model trained on surveillance data only, a chimeric model improved forecasts of weekly incident hospitalizations five weeks or more before the observed peak and for horizons greater than four weeks~(Fig.~\ref{fig.scoreweek} heatmap).
The chimeric model had poorer performance for short-term~(1-4 weeks) forecasts near the observed peak, but public health decision making at the peak has little impact and most public health policy changes take longer than 4 weeks to be enacted~\cite{dey2021lag}.
\begin{table}[ht!]
    \centering
    \begin{tabular}{r|lll}
      \hline
      Covariate & $\beta$ & (95CI) & p   \\
      \hline
      Forecast horizon        & -0.071 &(-0.073, -0.070) & ${<}0.01$\\
      Weeks from peak         &  \hspace{2px}0.044 &(0.040, 0.048)   & ${<}0.01$\\
      Horizon*Weeks from peak & -0.004 &(-0.004, -0.004)  & ${<}0.01$\\
      \hline
    \end{tabular}
    \caption{Linear regression fit to relative WIS (chimeric vs control) as dependent variable. Independent variables:~forecast horizon and the number of weeks at which the forecast was generated. $R^{2}=0.59$. \label{tab.regress}}
\end{table}
A linear regression was fit to RWIS as the dependent variable, and forecast horizon and the number of weeks before the peak~(where negative values indicate before the peak and positive values after the peak) as independent variables.
The regression table is presented in table~\ref{tab.regress}. 
Every week added to forecast horizon~(a longer forecast) has an estimated decrease~(i.e.~improvement) of 0.071 in RWIS.
However, generating forecasts one week closer to the true peak has an estimated increased~(i.e.~worsening) of 0.044 in RWIS. 
These regression results mimic Fig.~\ref{fig.scoreweek}A.

\subsection{In sample vs out of sample human judgment forecasts}
\label{result3}
\begin{figure}[ht!]
    \centering
    \includegraphics[width=1\linewidth]{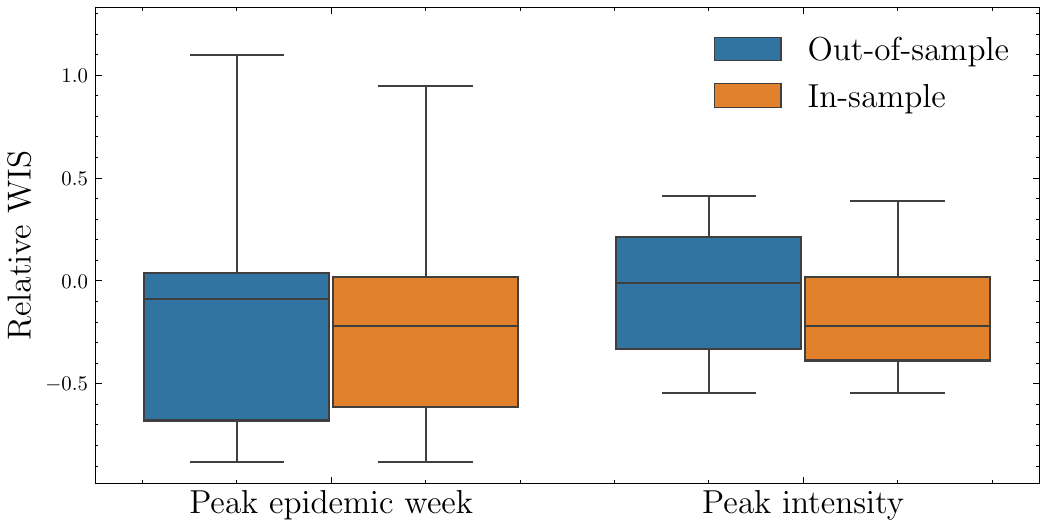}
    \caption{Relative WIS for human judgment-only ensemble forecasts of peak epidemic week and peak intensity over the influenza season stratified by: the ten states that were presented to humans~(orange, called in-sample forecasts) vs forecasts from a model that extended human judgment to the other 40 states~(blue, called out-of-sample, see Materials and Methods C. for the model).
    Forecasts are relative to the first forecasts generated at the beginning of the season.
    \label{fig.insample_vs_out} }
\end{figure}

The performance of human judgment forecasts of peak time were similar between states presented to humans~(`in-sample' forecasts) and states not presented to humans but generated by a model to extend human judgment~(`out-of-sample' forecasts; Mann-Whitney p=0.17).
For peak intensity, in-sample forecasts had better performance compared to out-of-sample forecasts~(p${<}$0.01). 
See Figure~\ref{fig.insample_vs_out}.

The performance of forecasts~(measured with RWIS) was relative to the first forecast generated at the beginning of the season. 
RWIS values were positive~(i.e.~worse than the original forecast generated at the start of the season) for both out and in-sample forecasts of both peak epidemic week and peak intensity~(See Table~\ref{tab.in_and_out}). 

\begin{table}[ht!]
    \centering
    \begin{tabular}{r|ll}
    \hline
        \multirow{2}{*}{Target} &  \multicolumn{2}{c}{Median RWIS (25th, 75th)} \\
               \cline{2-3}
               & In-sample & Out of sample \\
               \hline
         Peak epidemic week & -0.61 (-0.22, 0.02)  & -0.68 (-0.09, 0.04) \\
         Peak intensity     & -0.39 (-0.22, 0.02)  & -0.33 (-0.01, 0.21) \\ 
         \hline
    \end{tabular}
    \caption{ The Median, 25th, and 75th percentiles of relative WIS for in vs out of sample human judgment-only ensemble forecasts for forecasts of peak epidemic week and peak intensity. Forecasts are relative to the first forecast generated at the beginning of the season.}
    \label{tab.in_and_out}
\end{table}

\section{Discussion}

We presented 2 key findings that support the primary hypothesis of this study, that augmenting an epidemic model with human judgment can improve forecast performance: 
\begin{enumerate}[itemsep=-1px]
    \item Compared to a control model trained only on surveillance, a model trained on both surveillance and human judgment forecasts of peak epidemic week and peak intensity---a chimeric model---produced significantly better forecasts of peak epidemic week and significantly better forecasts of peak intensity. 

    \item For forecasts of incident hospitalizations longer than four weeks, the forecast performance of a chimeric model was better than control. 
\end{enumerate}

We also found that:~(1)~compared to the chimeric model, a HJ-only ensemble generated improved forecasts of peak epidemic week at the time of the observed peak~(though this true peak was not known to the model until the season ended) and four weeks after the true peak was observed; (2)~a control model, compared to chimeric, performed better for short-term forecasts from one to four weeks. 

Chimeric vs control forecast performance can be separated into three distinct cases:~(1)~chimeric short-term and long-term forecasts are improved compared to control; (2)~some chimeric short-term forecasts are poorer than control and others are improved while long-term forecasts are improved; and (3)~all chimeric short-term forecasts are poorer than control but long-term forecasts are improved. 

All three cases are related to how human judgment predictions influence the estimated number of susceptibles, causing the chimeric model to forecast smaller peaks and not assume explosive growth.
In case one, compared to the control, the median forecast is smaller than the observed number of incident hospitalizations. 
However, the 95\% prediction interval covers the true trajectory of hospitalizations and is much narrower than the control model prediction interval~(Fig.~\ref{fig.scoreweek}A.~orange vs blue).
In case two, the observed number of hospitalizations has explosive growth, peaks, and then slowly returns to zero hospitalizations. 
The chimeric model captures this period of explosive growth, but not as well as the control model which assumes tremendous growth in the number of hospitalizations.
This indicates that the control model assumes a much larger number of susceptibles exist in the system, making the chimeric model more accurate long-term~(Fig.~\ref{fig.scoreweek}B.)
Case three is an exaggerated instance of the previous case~(Fig.~\ref{fig.scoreweek}C.).
Here, again, the chimeric model's smaller assumed number of susceptibles leads to missing the period of explosive growth.
As a result, the control model has improved short-term forecasts. 
The chimeric model, though, makes superior long-term forecasts and does not `overshoot' the number of hospitalizations like the control model. 

At the state level, without data on case counts, there is an enormous number of potentially susceptible individuals compared to reported hospitalizations.
Without human judgment forecasts of the peak epiweek and peak intensity to lower the estimated potential number of susceptible individuals, the control model is more likely to assume a large number of susceptible individuals and thus, exponential growth of cases and hospitalizations.
A chimeric model, we assert, is a better model because it does not assume uncontrolled, explosive growth in hospitalizations. 

The improved longer-term forecasts generated by the proposed chimeric model allow public health officials and clinical staff more time to respond to upcoming changes in disease intensity~\cite{borchering2023public,howerton2023evaluation,howerton2023informing,loo2024us}.
In fact, public health officials have previously requested longer-term forecasts to aide in decision making~\cite{dey2021lag}, but to date, noise from surveillance systems has limited the majority of forecasting efforts to four weeks~\cite{drake2006limits,rosenkrantz2022fundamental,parag2022fundamental}.
The addition of human judgment forecasts of peak epiweek and peak intensity may be one path forward to generating the longer-term forecasts desired by public health officials. 

On their own, human judgment only ensemble forecasts of the peak epidemic week and peak intensity can be considered long-term forecasts~(the average peak was observed 13 weeks after the start of the season); and these forecasts, compared to control, showed improved performance.
Even though human judgment forecasts of the peak epidemic week were often later than the truth and forecasts of peak intensity tended to be larger than the truth, (See supplemental Figures~\ref{supp.hj_intes} and \ref{supp.hj_time}), the 95\% prediction interval for the majority of states included the true peak intensity and true peak epidemic week as far as 10 weeks before the peak was observed.
This demonstrates that a human judgment only ensemble alone can provide useful information for public health officials far enough ahead to impact decision making.
Coupling surveillance data and a human judgment ensemble allows practitioners to ask a richer set of questions about an epidemic~(weekly incidence in addition to peak epiweek and peak intensity) and receive accurate answers.

The results presented for the chimeric model depend on how to incorporate human judgment ensemble forecasts of peak epidemic week and peak intensity~(See Methods \ref{chimeric_model}), and the quality of human judgment forecasts themselves~(See Materials and Methods~\ref{data}~and~\ref{hjbuild}).

The copula that was used to build a joint density of human judgment ensemble forecasts of peak epidemic week and peak intensity assumed these two types of forecasts were independent from one another.
One improvement to the copula could be to estimate the correlation between observed peak epiweek and peak intensity and use this to inform a Gaussian copula.
In addition to the copula, we used a weighting function to determine how much human judgment forecasts, compared to surveillance, influenced model parameter estimates. 
In this work we choice the naive weight of one~(i.e. that surveillance and human judgment have the same weight). 
Because we observed that the chimeric model generates well-performing long-term forecasts and poorer short-term forecasts, a second improvement to the chimeric model could propose a weighting function that assigns large weight to surveillance data for forecast horizons within 4 weeks and then larger weight to human judgment for horizons greater than 4 weeks. 
Future improvements to the chimeric model such as optimizing the weighting function will only be as good as the quality of human judgment ensemble forecasts themselves.

To improve human judgment ensemble forecast performance, we recommend relying on an established human judgment platform to record and combine results.
Professional platforms have spent time standardizing the process of data storage and extracting unbiased forecasts from humans.
A standard set of data, visuals, and interactive platform to explore past incidence data and in particular past peaks could be created to support forecasts.
Recent work also suggests that a different elicitation technique---a method to extract from an individual a forecast---may be warranted when collecting human judgment predictions in service to a computational model~\cite{ibrahim2021eliciting}.
Importantly, the cognitive energy of experts who generate forecasts is limited.
The algorithm that we present in the methods to extend human judgment forecasts of ten states to all states is only one possible implementation and we expect future work can improve forecast results further.

This modeling approach may be particularly attractive when there is sparse data about a novel pathogen or when past surveillance of an outbreak in one location may not be readily applied to an outbreak in a new location.
For example, a chimeric approach that integrated human judgment and surveillance may have been useful during the 2022/23 mpox outbreak in non-endemic countries~\cite{mcandrew2022early,alcami2023pathogenesis}. 
In particular, modeling groups noted the difficulty with forecasting the peak for mpox in the US~\cite{charniga2024nowcasting}.

The presented chimeric modeling approach is scaleable to any public health reporting unit, any infectious agent that is recorded by a surveillance system, and to any spatial scale: national, health and human service region, state, county, city. 
The epidemic model for how cases and hospitalizations evolve over time and the algorithm to extend forecasts to all locations are separated so that the scientific community can further improve upon both efforts.

However, the `intuitive' approach of human judgment forecasting may make it difficult for public health officials to trust output from chimeric models.
Additional effort should be spent introducing public health officials to the benefits and drawbacks of human judgment as an analytical technique.

\subsection{Data}
\label{data}

The number of weekly incident hospitalizations~(hosps) due to influenza was collected from all states in the US beginning on August 5, 2023 and ending on April 27, 2024. 
Weeks are reported using the CDC standard `epidemic week'.
For this work, the first forecast was produced on Oct.~14, 2023. 

Metaculus forecasts were collected from individuals beginning October 18, 2023 and ending on March 2, 2024. 
Two questions were assigned for the state with the largest population in each of the 10 Health and Human Service regions (California, Texas, Florida, New York, Pennsylvania, Massachusetts, Illinois, Missouri, Colorado, and Washington) asked individuals to build a probability density over the week with (1) the maximum number of reported, incident hospitalizations~(peak epidemic week) and (2) on that week, the number of incident hospitalizations~(peak intensity).
Forecasters on Metaculus were allowed to submit an original forecast and then revise their original forecast as many times as they wanted within the influenza season. 

\subsection{Human judgment ensemble building}
\label{hjbuild}
Given a set of probability density functions from $N$ individuals, an ensemble probability density, $p$ is defined as  
\begin{align*}
    p(x) = \sum_{ j=1 }^{N} \pi_{j} p_{j}(x) \label{linearpoolhj}\\
    \forall_{j} \pi_{j} \geq 0;\; \sum_{j=1}^{N} \pi_{j} = 1
\end{align*}
where $p_{j}(x)$ is the density value that individual $j$ assigned to $x$, and $\pi_{j}$ is a weighting function that depends on an individual's past performance. 
The Metaculus ensemble that was used is a proprietary algorithm for assigning weights that depend on an individual's past performance. 

\subsection{Model to extend human judgment forecasts}
\label{hjextend}

The algorithm we propose to extend forecasts from ten states where we collected human judgment forecasts~(in-sample forecasts) to forecasts for all US states~(including `out-of-sample' states where we did not collect human judgment forecasts)~can be separated into two steps. 
\subsubsection{Step 1:~Model out-of-sample hosps as a convex combination of in-sample hosps}
We suppose that incident hospitalizations for the out-of-sample location at time $t$, $\underline{y_{t}}^{\text{oos}}$, is a weighted average of hospitalizations from the ten locations where human judgment forecasts were collected. 
\begin{align*}
    \underline{y_{t}}^{\text{oos}} \sim \text{Pois} \l( \sum_{k=1}^{10} w_{k} \underline{y_{t}}^{k}  \r) ; \; 
    \sum_{k=1}^{10} {w_{k}}=1; \; \forall_{k} w_{k}\geq 0
\end{align*}
where $ \underline{y_{t}}^{k}$ is the \textit{scaled} incident hospitalizations for in-sample location $k$ at week $t$, and $w_{k}$ is a non-negative weight assigned to location $k$.
Weights were fitted by minimizing the negative log-likelihood of past surveillance data~\cite{blank2020pymoo}.

We scaled hospitalizations  to be between zero and one by using each state's historical number of incident hospitalizations so that the above combination is not biased by population size for the ten states with in-sample forecasts.
Other options to scale, such as scaling by population size, could be used.
To scale incident hospitalizations, the number of incident hospitalizations for location $k$ up until week $t$ is divided by the maximum observed hospitalizations for this same location up until the most recently observed number of hospitalizations at time $t$ which we call $c_{k}$.

The output from this step is a vector of ten fitted weights or $\hat{w} = [w_{1},w_{2},\cdots,w_{10}]$ for all US states.

\subsubsection{Step 2:~Build a mixture model using fitted weights}
\begin{figure}[ht!]
    \centering
    \includegraphics[width=1.0\columnwidth]{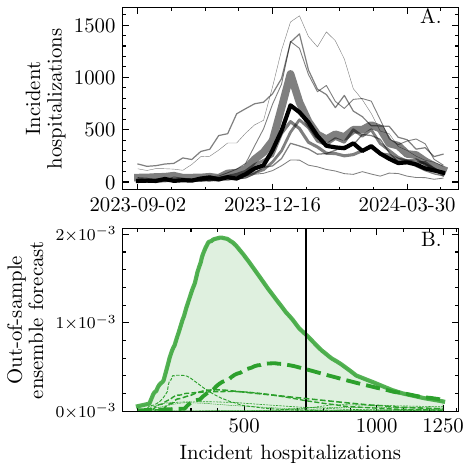}
    \caption{An example extending human judgment forecasts spatially, from ten `in-sample' states where human forecasts were collected to an `out-of-sample' state~(NJ) where human forecasts were not collected.
    (A.)~Step one estimates weights for in-sample states by modeling incident hospitalizations~(hosps) for the out-of-sample state~(black solid line) as a convex combination of in-sample hosps~(ten gray lines; thicker lines indicate larger weights). 
    (B.)~Step two takes as input weights from step one and in-sample human forecasts~(dashed lines, where thicker dashes indicate larger weights) to 
    build a mixture model for NJ~(solid green line and shading).  
    True peak hosps is denoted as a black vertical line.
    \label{fig.alg}}
\end{figure}

Given as input a set of `in-sample' forecasts, represented as probability densities, for ten locations $f = \{f_{1},f_{2},\cdots,f_{10}\}$ and associated weight vector~$\hat{w}$ from step 1, we assign a forecast $z$ for location $L$ as 
\begin{align*}
    g(y) &= \sum_{k=1}^{10} \hat{w}_{k} c_{k} f_{k}(y \cdot c_{k}); y \in [0,1]\\ 
    z(x) &= c_{L}^{-1} \,g(x/c_{L} )
\end{align*}
where $g$ is a compound density for location $L$ with $[0{-}1]$ support by scaling each human judgment forecast $f_{k}$ by $c_{k}$, and $z$ is the density $g$ placed back on a natural scale for incident hospitalizations that occur in state $L$.
See Fig.~\ref{fig.alg} for an example.  
%
\subsection{Control model}
\label{control_model}
As a control model, we fit to observed incident hospitalizations~($o_{1},o_{2},\cdots,o_{t}$) a discrete time Kermack-McKendrick model with compartments for susceptible~($S$), infected~($i$), and hospitalized~($h$) disease states~\cite{diekmann2021discrete}. 
The latent number of individuals in each disease state evolves according to 
\begin{align*}
    \text{FOI} &= \exp\left\{ - \lambda(t) \sum_{j=1}^{\text{LAG}}  i_{t-j} a(j;\beta_{g}) \right\} \\
    S_{t} &=  (\text{FOI})  S_{t-1} ; \;
    i_{t} = (1-\text{FOI}) S_{t-1}  \\
    h_{t} &=  \sum_{j=1}^{\text{LAG}} i_{t-j+1} q(j;\beta_{e}) \\ 
    S^{0}(t) &=  N \alpha  - \left[i^{0}(t) + h^{0}(t)\right] \\ 
    i^{0}(t) &=  N  \exp( \mu_{\text{infections}} t ) ;\;
    h^{0}(t) =  N  \exp( \mu_{\text{hosps}} t )
\end{align*}
where $a(t)$ is the distribution of the serial interval and $q(j)$ is the lag between an incident infection and hospitalization. We chose the Poisson distribution for both $a$ and $e$. 
The parameters $\mu_{\text{infections}}$ and $\mu_{\text{hosps}}$ control the rate of incident infections and hospitalizations before the observational period~(i.e., ~before the first recorded number of incident hospitalizations), and $\alpha$ controls the number of individuals in the system who are initially susceptible to disease.
$\text{LAG}$ is the number of time units to consider in the force of infection~(i.e.~the rate at which susceptibles are infected) and is an approximation~(See~\cite{diekmann2021discrete} for details).
The time-dependent $\lambda(t)$ that controls the force of infection $\text{FOI}$ is parameterized as 
\begin{align*}
    \lambda(t; R, \mu, \sigma, s^{2}) &= R \cdot b(t) + \epsilon_{t}\\
    b(t) &= \exp \left \{ -\frac{1}{2}\left[(t - \mu)/\sigma\right]^{2}\right\}\\ 
    \epsilon_{t} &\sim \sum_{t=1}^{t} Z_{t}; \; Z_{t} \sim \mathcal{N}(0,s^{2})
\end{align*}
where $b(t)$ takes the form of a Gaussian density normalized to equal one when $t=\mu$, and $\epsilon_{t}$ is a Gaussian random walk~\cite{edlund2011comparing}. 
We assume incident hospitalisations follows a Poisson observation process,
$
    o_{t} \sim \text{Pois}(h_{t}). 
$
~We take a Bayesian approach to fit the model by assuming prior densities over parameters as 
\begin{align*}
    \beta_{g} \sim \text{Gamma}(1,1); \; \beta_{e} \sim \text{Gamma}(1,1) \\ 
    \mu_{\text{infections}} \sim \text{Beta}(0.5,0.5) ; \;   
    \mu_{\text{hosps}}\sim \text{Beta}(0.5,0.5) \\ 
    \alpha \sim \text{Beta}(0.5,0.5) ; \; R \sim \text{Gamma}(1,1) \\ 
    \mu \sim \text{Unif}(0,T) ;\;
    \sigma \sim \text{Unif}(0,T); \; s \sim \text{HalfCauchy}(1)
\end{align*}
and use a No-U-Turn MCMC sampler to fit the model to weekly observed incident hospitalizations~\cite{phan2019composable,hoffman2014no}.
The posterior density over all parameters~($\theta$) is 
\begin{align*}
    p(\theta | \mathcal{D}) \propto \underbrace{p(\theta)}_{\text{Prior}} \times \overbrace{ \prod_{t=1}^{T} \text{Pois}(o_{t} | h_{t}) }^{\text{Data influence}}.
\end{align*}
We note that the model is run on a daily time scale, and incident hospitalizations are summed up to the time scale of a week. 
The MCMC sampler is given initial parameter values by first fitting the above model using a maximum likelihood approach and a genetic algorithm to search a feasible parameter space~\cite{blank2020pymoo}.

\subsection{Chimeric model}
\label{chimeric_model}
The chimeric model is specified the same as the above control model except for the addition of human judgment ensemble probability density functions---one for peak epiweek~($t*$) and one for peak intensity~($h*$)---that influence the posterior density.
Given a location, we retrieve two vectors of human judgment predictive quantiles: one for the peak epiweek and a second for peak intensity. 
A cumulative density can be estimated via interpolation with monotonic cubic splines and probability density by numerical differentiation~\cite{interpax}.
A copula can be used to infer the joint density over peak epiweek~($f_{\text{peak epiweek}}$) and peak intensity~($f_{\text{peak intensity}}$) as, 
\begin{align*}
    f(t,h) =  c[ F_{\text{peak epiweek}}(t), F_{\text{peak intensity}}(h) ],
\end{align*}
where $c$ is a 2-dimensional joint probability density corresponding to  a joint cumulative density $C$~\cite{jaworski2010copula}. 
Though there likely exists a correlation between human judgment forecasts of peak epiweek and peak intensity, we chose the `independence' copula which considers peak epiweek and peak intensity as independent from one another, 
\begin{align*}
    f(t,h) =  f_{\text{peak epiweek}}(t) \cdot f_{\text{peak intensity}}(h).
\end{align*}
A copula that better captures the correlation between peak epiweek and peak intensity, we suspect, would perform even better than the independence copula~\cite{han2019mid,ray2017infectious,jeng2023application}.

To incorporate human judgment forecasts into the control model, we must
(1)~augment our parameter space to include peak epiweek and peak intensity as random variables~(See \cite{osthus2017forecasting} for a similar approach using seasonal data) and (2)~evaluate the model's proposed peak epiweek~$(\tau)$ and peak intensity~($\rho$) against our corresponding human judgment forecasts.
We include a human judgment term into the posterior density as 
\begin{align*}
    p(\theta| \mathcal{D}, \mathcal{H}) \propto \underbrace{p(\theta)}_{\text{Prior}} \times \overbrace{ \prod_{t=1}^{T} \text{Pois}(o_{t} | h_{t}) }^{\text{Data influence}} \times \\ \underbrace{ \exp\left\{\omega(t,x;\theta)\right\} \cdot c\left[F_{\text{peak epiweek}}, F_{\text{peak intensity}}\right] (\tau,\rho) }_{\text{Human judgment influence}}
    \end{align*}
where we include a weighting function $\omega(t,x;\psi)$ that can modify the influence of human judgment on the posterior density. 
This function depends on the time in the season $t$---because we expect that human judgment forecasts should improve as more surveillance data is observed---and potentially covariate information $x$ with associated covariates $\psi$.  
Covariate information could include influenza A vs B dominance, a H1 vs H3 dominant epidemic, or vaccine levels which may influenza when flu peaks and the number of hospitalizations as a result. 
For this work, we chose the independence copula~(as above) and the naive weight of $\omega(t,x;\theta)=1$.
We expect that a more realistic choice of copula and weighting function would lead to better results.

\begin{figure}
    \centering
    \includegraphics{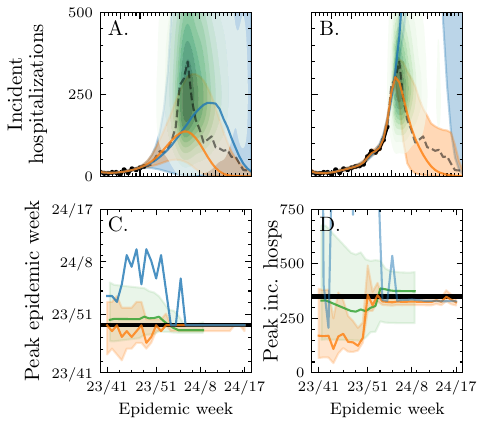}
    \caption{A chimeric~(orange) and control~(blue) model forecast represented as a median plus 80\% prediction interval plus human judgment joint density of peak epiweek and peak intensity~(green) for the state of Alabama with surveillance data up to Oct. 21, 2023~(A.) and Dec.~30, 2023~(B.).
    Incident hospitalizations are denoted by black circles and the remaining, unknown, incident hospitalizations are denoted by a gray dashed line.
    The median plus 80\% PI for the peak epidemic week~(C.) and peak intensity~(D.) for chimeric and control models plus human judgment ensemble.
    For control, only the median is presented. 
    Note that humans did not build forecasts for Alabama, instead forecasts of peak epiweek and intensity are generated from the algorithm in section~\ref{hjextend}. }
    \label{fig.chimericexample}
\end{figure}

\subsection{Forecast evaluation}
\label{evaluation}
We evaluated forecasts with the weighted interval score~(WIS)~\cite{bracher2021evaluating,bosse2023scoring}.
The WIS inputs a forecast designated as a vector of quantiles and outputs a non-negative score where smaller values indicate a better forecast. 
Given forecasts $f$ and $g$, to compare two models, we computed the relative WIS as 
\begin{align*}
    \text{RWIS}(f,g) = \frac{\text{WIS}(f)}{\text{WIS}(g)} - 1.
\end{align*}
where negative $\text{RWIS}$ indicate forecast $f$ is better than $g$ and vice versa.

A one-sample t-test was used to compare a difference in mean WIS between either the chimeric or human judgment only ensemble versus the control. 
A two-sample t-test was used to compare performance between chimeric and human judgment only ensemble forecasts.

\section{Acknowledgements}
This work has been supported with funding by Cooperative Agreement number (NU38OT000297) from the CDC and the Council for State and Territorial Epidemiologists (CSTE); award number (R35GM146974) from the National Institute of General Medical Sciences, National Institutes of Health (NIH); and award numbers (SES2200228 and IIS2229881) from the National Science Foundation (NSF). The content is solely the responsibility of the authors and does not necessarily represent the official views of CDC, CSTE, NIH, or NSF. Special thanks to Ryan Beck from Metaculus for support.



%

\clearpage
\appendix
\renewcommand{\figurename}{Supp.~Fig.}
\setcounter{figure}{0}

\begin{figure*}
    \centering
    \includegraphics[width=1\textwidth]{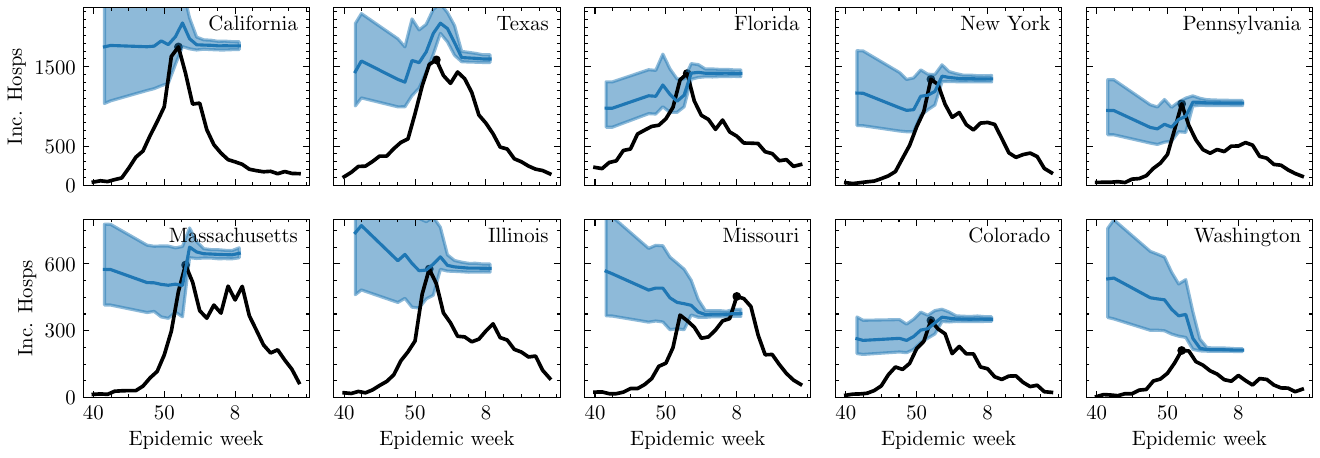}
    \caption{ Human judgment ensemble forecasts over the influenza season of peak intensity for the ten states with the largest population in each health and human service region. 
    Forecasts are represented as a median number of incident hospitalizations~(solid blue) and 95\% prediction interval~(blue shaded region). The observed number of hospitalizations is displayed in black and the peak intensity is denoted with a black circle. \label{supp.hj_intes}}
\end{figure*}

\begin{figure*}
    \centering
    \includegraphics[width=1\textwidth]{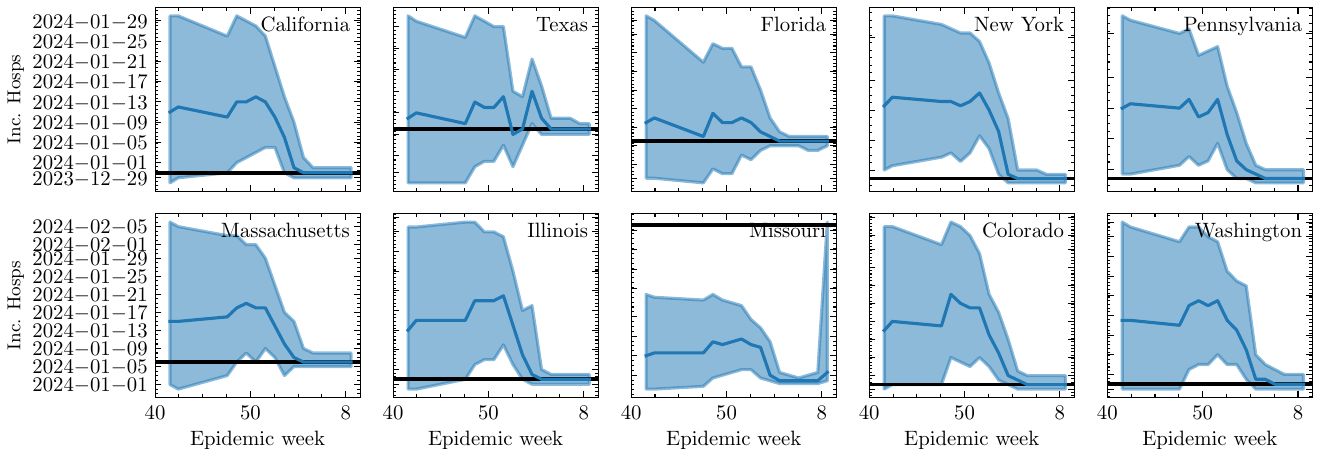}
    \caption{Human judgment ensemble forecasts over the influenza season of the peak epidemic week for the ten states with the largest population in each health and human service region. Forecasts are represented as a median week~(solid blue) and 95\% prediction interval~(blue shaded region). 
    The observed peak epidemic week is displayed as a black horizontal line. \label{supp.hj_time}}
\end{figure*}

\end{document}